%% file: main.tex
\documentclass[sigconf]{acmart}
\input{macros}

\AtBeginDocument{%
  }

\setcopyright{acmcopyright}
\copyrightyear{2024}
\acmYear{2024}
\acmDOI{XXXXXXX.XXXXXXX}
\acmPrice{15.00}
\acmISBN{978-1-4503-XXXX-X/18/06}

\begin{document}

\title[]{{\vibeC}: The Missing Quality Assurance Piece in Vibe Coding}

\author{Song Wang}
\email{wangsong@yorku.ca}
\affiliation{%
  \institution{York University}
  \city{Toronto}
  \state{ON}
  \country{Canada}
}

\input{sections/abstract}


\received{28 September 2023}
\received[revised]{5 March 2024}
\received[accepted]{16 April 2024}

\maketitle

\input{sections/intro}
\input{sections/background}
\input{sections/strategies}
\input{sections/directions}
\input{sections/future}
\input{sections/conclusion}

\bibliographystyle{ACM-Reference-Format}
\bibliography{sample-acmsmall-conf}


\end{document}

%% file: macros.tex
\usepackage{soul}
\usepackage{hyperref}
\usepackage{multirow}
\usepackage{listings}
\usepackage{fancybox}
\usepackage{enumitem}
\usepackage{graphicx}
\usepackage{color}
\usepackage{array}
\usepackage{amsmath}
\usepackage{xspace}
\usepackage{caption}
\usepackage{pgfplots}
\usepackage{balance}
\usepackage{subcaption}
\pgfplotsset{compat=1.16}
\usepackage{pgf-pie}
\usetikzlibrary{pgfplots.statistics,calc}
\usepackage{algorithm}
\usepackage{algorithmic}
\usepackage{adjustbox}
\usepackage{makecell}

\usepackage{tikz}
\newcommand*\circled[1]{\tikz[baseline=(char.base)]{
            \node[shape=circle,draw,inner sep=2pt] (char) {#1};}}

\usepackage{tcolorbox}
\makeatletter
\newcommand{\mybox}[1]{%
	\setbox0=\hbox{#1}%
	\setlength{\@tempdima}{\dimexpr\wd0+13pt}%
	\begin{tcolorbox}[boxrule=0.5pt, colback=white, arc=4pt,
		left=6pt,right=6pt,top=6pt,bottom=6pt,boxsep=0pt]
		#1
	\end{tcolorbox}
}

\definecolor{songcolor}{RGB}{191,191,191}

\newcommand{\vibeC}{\textit{VibeContract}}

%% file: sections/abstract.tex
\begin{abstract} 
Recent advances in large language models (LLMs) have given rise to vibe coding, a style of software development where developers rely on AI coding assistants to generate, modify, and refactor code using natural language instructions. While this paradigm accelerates software development and lowers barriers to entry, it introduces new challenges for quality assurance (QA). 
AI-generated code can appear correct but often contains hidden logical errors, creating an urgent need for novel QA approaches. 

In this vision paper, we propose the {\vibeC} paradigm as a missing piece in vibe coding. In this approach, high-level natural-language intent is decomposed into explicit task sequences, and task-level contracts are generated to capture expected inputs, outputs, constraints, and behavioral properties. Developers validate these contracts, and traceability is maintained between tasks, contracts, and generated code. Contracts then guide LLMs for testing, runtime verification, and debugging, enabling QA to occur continuously and proactively alongside code generation.

We demonstrate how the {\vibeC} paradigm can substantially improve the correctness, robustness, and maintainability of LLM-generated code through an example project. We argue that {\vibeC} introduces a structured and verifiable development workflow that transforms vibe coding from a fast but error-prone practice into a predictable, auditable, and trustworthy software engineering process.
\end{abstract}

%% file: sections/intro.tex
\section{Introduction}
\label{sec:intro}
\input{figure/intent-driven}

The rise of large language models (LLMs) has fundamentally transformed software development, giving birth to a new paradigm often referred to as vibe coding~\cite{vibeCoding}. In this approach, developers interact with AI assistants through natural language instructions, which generate, refactor, and optimize code across diverse tasks. Vibe coding promises significant benefits such as faster prototyping, reduced manual effort, and access to advanced coding capabilities for developers with varying levels of expertise~\cite{vibeCodingbenefit, pimenova2025good}. However, alongside these advantages, it introduces profound challenges for quality assurance (QA)~\cite{fawzy2025vibe}. Developers frequently encounter situations where the AI produces code that looks syntactically correct but is semantically flawed, leading to subtle bugs, degraded reliability, or security vulnerabilities~\cite{gao2025survey}. This phenomenon exemplifies an illusion of correctness, highlighting the urgent need for rigorous and systematic quality assurance approaches tailored to AI-assisted development. 

Existing strategies for ensuring correctness in AI-generated code are largely post-hoc, reactive, and often rely heavily on human inspection. In practice, developers are forced to manually validate AI outputs, reason about the AI’s hidden decision-making, and debug non-deterministic behaviors. Such practices are not scalable, and they undermine one of vibe coding’s primary benefits: accelerated software development~\cite{reviewAI}.

To address these challenges, we introduce the {\vibeC} paradigm, which integrates code generation with quality assurance by anchoring AI-generated code to explicit contracts, generated by AI and lightly verified by developers. 
As illustrated in Figure~\ref{fig:intentContract}, conventional Vibe Coding typically involves only Steps~\circled{1} (LLM-based intent reasoning from natural-language inputs) and~\circled{3} (LLM-based code generation). The {\vibeC} paradigm augments this workflow by introducing two additional steps, i.e., Steps~\circled{2} and~\circled{4}, which focus on contract generation and validation, and contract-based testing, respectively. In addition, {\vibeC} enhances the original LLM-based code generation in vibe coding (i.e., Step~\circled{3}) by incorporating contract-based guidelines that constrain and guide code generation. 
At its core, this paradigm treats AI-generated code as an implementation of explicit, validated task-level contracts derived from developer intent. By embedding quality assurance directly into the code generation workflow, the {\vibeC} paradigm transforms vibe coding from an exploratory and often error-prone process into a predictable, verifiable, and reliable development workflow. It provides a structured mechanism for mitigating the risks associated with the non-deterministic nature of LLM-based code generation, while preserving developer productivity and improving correctness.

To illustrate the effectiveness of the {\vibeC} paradigm, we use a representative project involving the implementation of an Automated Teller Machine (ATM) system in Java. This example captures a range of common software engineering challenges, including managing multiple interdependent classes, enforcing domain-specific constraints (such as account balances and transaction limits), and handling user inputs robustly. By applying {\vibeC}, we demonstrate how high-level natural-language prompts can be decomposed into discrete tasks, each accompanied by precise contracts that specify inputs, outputs, and expected behaviors. The contracts guide code generation, verification, and testing, enabling the LLM to produce reliable code. 

In this paper, we outline our research vision on integrating program contracts into AI-assisted software development, summarizing our contributions as follows: 

\begin{itemize}

\item We introduce the {\vibeC} paradigm, a novel workflow that integrates code generation with quality assurance by anchoring AI-generated code to explicit, developer-verified contracts. 

\item We derive concrete research goals, opportunities, and challenges, outlining directions for advancing contract-driven AI-assisted software development to improve correctness, robustness, and developer productivity.

\end{itemize}

%% file: figure/intent-driven.tex
\begin{figure}[t!]
  \centering  \includegraphics[width=0.5\textwidth]{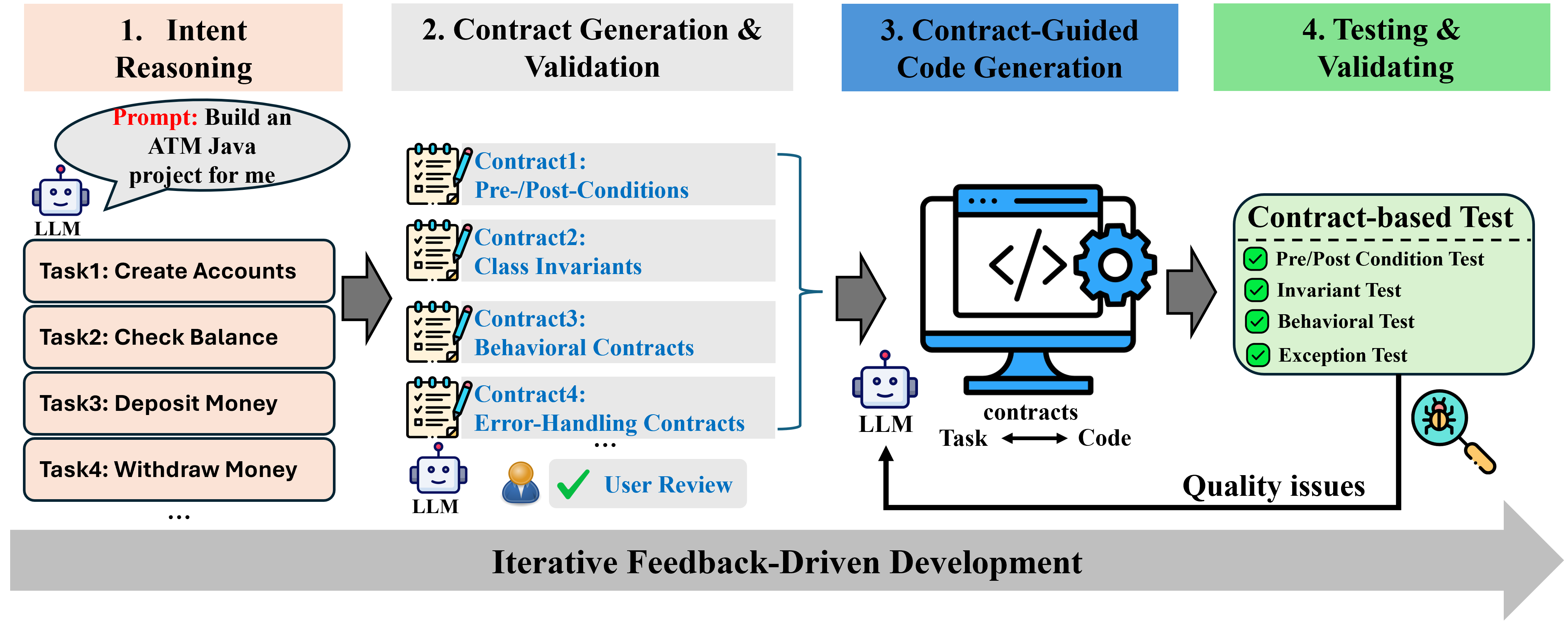}
  \caption{An overview of the {\vibeC} paradigm for Vibe Coding. Notably, the original Vibe Coding approach included only Steps 1 and 3.}
  \label{fig:intentContract}
\end{figure}

%% file: sections/background.tex
\section{Background and Related Work}
\label{sec:background}

\subsection{Vibe Coding}
\label{sec:vibe}

Vibe Coding is a paradigm for LLM-assisted code generation that leverages LLMs to transform high-level natural-language prompts into executable code~\cite{sapkota2025vibe,vibeCoding,ge2025survey}. The core idea is to bridge the gap between human intent and code implementation by using LLMs to interpret, decompose, and synthesize software artifacts. Unlike traditional code generation approaches, which rely on static templates or formal specifications, Vibe Coding emphasizes iterative, task-oriented generation where high-level instructions are broken down into granular, actionable units, enabling step-by-step reasoning and code synthesis~\cite{vibeCodingbenefit}.


Vibe Coding has shown promise in rapidly prototyping software, automating repetitive coding tasks, and enabling non-experts to generate working programs from natural-language instructions. However, it also faces challenges, particularly in ensuring correctness, handling subtle domain constraints, and maintaining traceability~\cite{ray2025review,fawzy2025vibe}. These challenges motivate us to propose {\vibeC}, which integrates formal contracts into the Vibe Coding workflow to improve reliability, enforce constraints, and make the generated code verifiable and maintainable. 

To improve the quality of Vibe Coding, recently, Speckit~\cite{speckit} was proposed, which represents spec-driven development for LLM-based software engineering by structuring development around intent clarification, specification refinement, planning, and code generation, with specifications as first-class artifacts. Our contract-based paradigm is complementary: it can integrate with specifications generated by Speckit and transform them into explicit intent contracts, strengthening constraint enforcement and behavioral guarantees in LLM-driven development.

\subsection{Design by Contract (DbC)}
\label{sec:dbc}

Design by Contract (DbC) was formally introduced by Meyer as a disciplined approach to software construction based on explicit, executable specifications of component behavior~\cite{meyer2002applying}. In DbC, software components interact through well-defined contracts consisting of preconditions, postconditions, and invariants. These contracts precisely define the obligations of callers and the guarantees provided by callees, enabling correctness by construction and early detection of logical errors. \textbf{Preconditions} specify the conditions that must hold before a method or operation is invoked, \textbf{postconditions} define the conditions that must hold after successful execution, and \textbf{invariants} capture properties that must always remain true for a component or system throughout its lifetime. 

The practical adoption of DbC has been supported by specification languages and tooling. Eiffel provides native support for contracts as executable specifications and runtime checks~\cite{meyer1992eiffel,paige2004specification}, while the Java Modeling Language (JML) enables contract annotations for Java programs with static and dynamic verification~\cite{leavens2006design}. Empirical studies by Schiller et al. showed that contracts can uncover subtle bugs and specification mismatches, but also raise usability and scalability challenges~\cite{schiller2014case}. 
As ML systems became integral to modern software, DbC principles were extended to data-driven components. Ahmed et al. introduced contracts over data shapes, value ranges, and semantic assumptions for deep learning APIs, revealing that many ML failures stem from violated data assumptions~\cite{ahmed2023design,ahmed2024inferring}. Related work, such as MLGuard and contract-based architectural validation, further demonstrates the relevance of contracts for ensuring robustness and trustworthiness in ML-enabled systems~\cite{wong2023mlguard,meijer2024contract}.

In this paper, we extend Vibe Coding by incorporating DbC principles, enabling the automatic generation of code with explicit, verifiable contracts. By integrating contracts into the Vibe Coding workflow, we aim to improve code correctness, detect subtle logic errors early, and provide developers with formal specifications alongside generated code. 

%% file: sections/strategies.tex
\section{{\vibeC} Approach}
\label{sec:strategires}

Figure~\ref{fig:intentContract} illustrates our vision of {\vibeC}, a framework designed to bridge high-level developer intent with reliable, contract-driven code generation. The approach consists of four key steps:

\begin{itemize}

\item  \textbf{Intent Decomposition:} This is the original initial step of Vibe Coding, in which high-level natural-language prompts are analyzed and decomposed into a sequence of executable tasks. Often, a Chain-of-Thought~\cite{wei2022chain} or thinking LLM model is employed to systematically interpret the prompt. Each task represents a granular, actionable unit of work, allowing the system to capture complex intent while preserving semantic meaning. This decomposition forms the foundation for precise contract specification. 

\item  \textbf{Contract Generation \& Validation:} As introduced in the Section~\ref{sec:intro}, this step enhances the original Vibe Coding workflow. For each decomposed task, {\vibeC} uses an LLM to generate a set of formal contracts specifying inputs, outputs, constraints, and expected behaviors. Developers then review and validate these contracts to ensure they accurately reflect the intended functionality. This process produces a clear, machine-readable specification that serves both as documentation and as a correctness safeguard. 

\item  \textbf{Contract-Guided Code Generation:} This step extends the original Vibe Coding process by leveraging validated contracts to guide code generation. In {\vibeC}, the LLM generates code under explicit contract-based guidelines that specify expected inputs, outputs, invariants, and behavioral constraints. These contracts act as structured supervision for the generation process, reducing ambiguity in natural-language prompts and constraining the solution space. 

\item  \textbf{Contract-Guided Testing \& Verification:} This step leverages the previously generated contracts to systematically guide the creation of targeted tests. By encoding the expected inputs, outputs, and behavioral constraints, contracts provide a precise reference against which the generated code can be validated. Any detected logic errors, contract violations, or mismatches between the code and the intended behavior are automatically reported and fed back into the code generation process. This feedback loop enables iterative refinement, ensuring that subsequent code generation aligns more closely with the original intent and contracts, thereby improving correctness and maintainability throughout the development workflow.

\end{itemize}

Together, these steps form a closed-loop system in which high-level intent is systematically translated into validated and verifiable code. By integrating contract reasoning at every stage, {\vibeC} enhances code reliability, supports developer oversight, and lays the groundwork for more trustworthy automated code generation.

%% file: sections/directions.tex
\section{Illustration Example}
\label{sec:directions}

In this section, we use an ATM system as a running example to illustrate how {\vibeC} operates and improves the quality of code generated by LLMs. Specifically, we begin by following the conventional Vibe Coding workflow to generate the code and highlight its quality issues. We then apply {\vibeC} to regenerate the code, demonstrating how the proposed paradigm addresses and mitigates these issues. The initial prompt used for code generation in both runs is ``\textit{Build an ATM Java project for me}'', and we employ GPT-5.2 Instant\footnote{\url{https://openai.com/index/introducing-gpt-5-2/}}
 as the LLM to drive the code generation process.
\input{figure/code-structure}

\subsection{Quality Issues in Code Generated by Conventional Vibe Coding}
\input{figure/original}
Following the original Vibe Coding workflow, the LLM first produces a preliminary implementation based on the given prompts, i.e., ``\textit{Build an ATM Java project for me}''. The resulting project structure, shown in Figure~\ref{fig:atm}, consists of five Java classes: \textit{ATM}, \textit{Account}, \textit{Bank}, and a driver class \textit{Main}. The complete project code is available in our replication package\footnote{https://zenodo.org/records/18136814}.

We use the \textit{Account} class as a running example to highlight quality issues in code generated by original Vibe Coding. As shown in Figure~\ref{fig:original}, the original (simplified, as we only kept the constructor to highlight the quality issue) Account class was generated by GPT-5.2 Instant using the standard Vibe Coding workflow.
As we can see, there are critical logic flaws in this generated code. Upon initialization with the constructor method, an \textit{Account} object’s \textit{balance} should be non-negative and finite, and its \textit{accountNumber} and \textit{pin} should be valid. However, the generated implementation does not enforce these constraints, leaving the system vulnerable to invalid or unexpected states. This example illustrates a broader challenge with automated code generation, i.e., subtle but important domain-specific rules are often overlooked, which can compromise correctness and reliability. 

\subsection{How {\vibeC} Addresses Code Quality Issues}

\input{table/contracts}

We further apply {\vibeC} to regenerate the code using the same prompt and the same LLM (GPT‑5.2 Instant). 
To save space, we show only the constructor of the \textit{Account} class. 
The contracts generated by the LLM are shown in Table~\ref{Tab:contracts}. As we can see, the LLM generates a comprehensive set of contracts regarding the input and output constraints of the constructor. In particular, it produces four preconditions corresponding to different input parameters and four postconditions, such as $\{balance >= 0\}$ (i.e., the balance of a newly created account should be non-negative) and $\{!Double.isNaN(balance)$\&\& $!Double.isInfinite(balance)\}$ (i.e., the balance of a newly created account should not be either $NaN$ or $Infinite$). We have manually verified all of these contracts, confirming their correctness. Based on the generated contracts, we instruct the LLM to reimplement the \textit{Account} class. As shown in Figure~\ref{fig:contractenhanced}, explicitly guiding code generation with these contracts, particularly those related to the \textit{balance}, \textit{accountNumber}, and \textit{pin}, successfully corrects the logic errors present in the original vibe coding output (Figure~\ref{fig:original}). For example, with respect to \textit{balance}, the LLM adheres to the specified contract by enforcing its constraints and providing appropriate error handling when an invalid \textbf{balance} value is supplied. 
This contract-guided approach ensures that all critical constraints are properly enforced, resulting in a more robust and reliable implementation that faithfully aligns with the developer’s intended behavior. 

This example demonstrates that {\vibeC} can help capture the developer’s intent and enforce critical constraints that were previously overlooked in the original Vibe Coding. By explicitly incorporating these contracts, the regenerated code becomes more robust and better aligned with domain-specific requirements.

\input{figure/contract-enhanced}

%% file: figure/code-structure.tex
\begin{figure}[t!]
  \centering
  \includegraphics[width=0.2\textwidth]{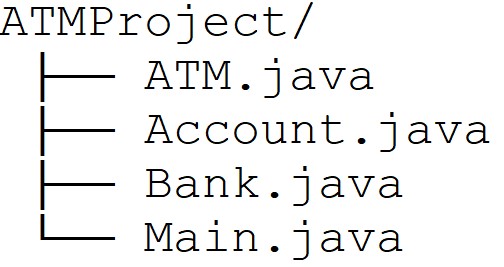}
  \caption{The structure of the ATM project generated by GPT-5.2 Instant.}
  \label{fig:atm}
\end{figure}

%% file: figure/original.tex
\begin{figure}[t!]
  \centering
  \includegraphics[width=0.48\textwidth]{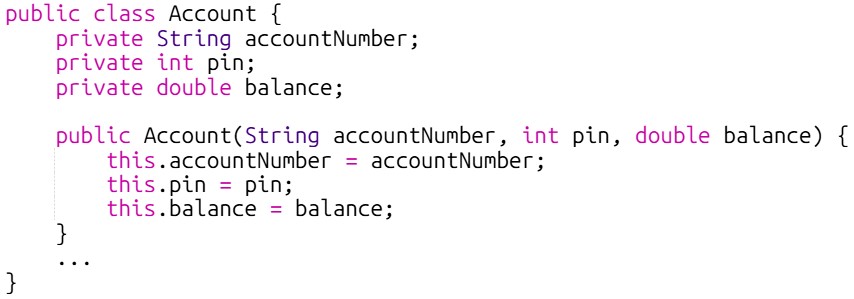}
  \vspace{-0.1in}
  \caption{The original (simplified) \textsc{Account} class generated by GPT-5.2 Instant via vibe coding. There are several logic flaws, i.e., when an \textsc{Account} object is initialized, the  \textit{balance} should be non-negative and finite, the  \textit{accountNumber} and \textit{pin} should be valid. However, the generated code fails to account for these subtle yet important constraints.}
  \label{fig:original}
\end{figure}

%% file: table/contracts.tex
\begin{table}[t!]
\centering
\caption{Contracts generated by the LLM for the parameters of the \textit{Account} class constructor.}
\resizebox{0.5\textwidth}{!}{
\begin{tabular}{|l|l|l|}
\hline
\textbf{Contract type} & \textbf{Element} & \textbf{Contract} \\ \hline
Precondition  &  accountNumber      &    $accountNumber != null$ \&\& $!accountNumber.isEmpty()$      \\ \hline
Precondition  &  pin    &  $0<= pin$ \&\& $pin <= 9999$     \\ \hline
Precondition  &  balance    &   $this.balance >= 0$       \\ \hline
Precondition  &  balance     &  $!Double.isNaN(balance)$\&\& $!Double.isInfinite(balance)$        \\ \hline
Postcondition &  accountNumber  &  $this.accountNumber == accountNumber$        \\ \hline
Postcondition &  pin    &   $this.pin == pin$      \\ \hline
Postcondition &  balance   &   $this.balance == balance$       \\ \hline
Postcondition &  balance  &   $this.balance>=0$      \\ \hline
\end{tabular}
}
\label{Tab:contracts}
\end{table}

%% file: figure/contract-enhanced.tex
\begin{figure}[t!]
  \centering
  \includegraphics[width=0.45\textwidth]{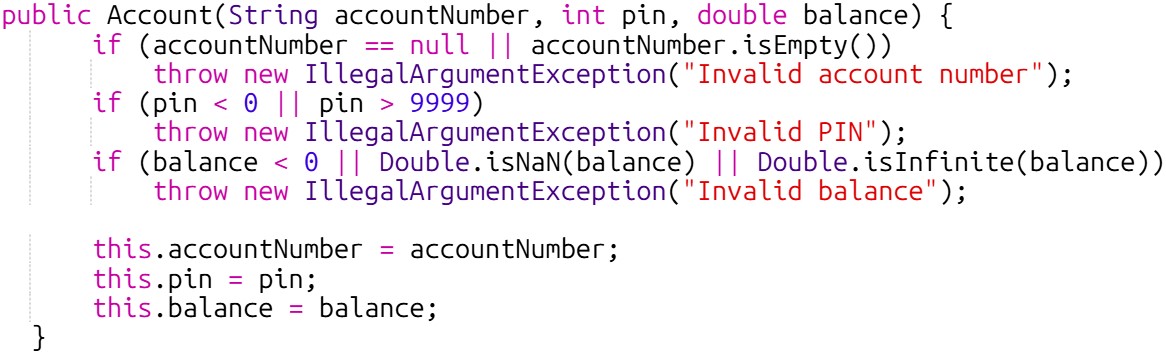}
  \vspace{-0.1in}
  \caption{Code generated by {\vibeC} with contracts applied.}
  \label{fig:contractenhanced}
\end{figure}

%% file: sections/future.tex
\section{Future Directions}
\label{sec:fplans}

While the {\vibeC} paradigm demonstrates the benefits of contract-guided Vibe Coding, it also opens up a range of new research questions. 
We outline three promising future directions that build on our findings for advancing contract-driven Vibe Coding. 

\subsection{Automated Contract Synthesis and Refinement}

One of the key limitations of current {\vibeC} workflows is the reliance on manual verification of contracts. Future research could focus on automating contract synthesis and refinement using a combination of static analysis, symbolic execution, and LLM reasoning. By analyzing both the natural-language prompt and existing code patterns, the system could automatically propose preconditions, postconditions, and invariants. Additionally, contract refinement can be performed iteratively by analyzing runtime violations or test failures, feeding these back into the model to improve future contract suggestions. For instance, in the \textit{Account} class example, the system could automatically detect that $balance >= 0$ and $!Double.isNaN(balance)$ are necessary preconditions based on code usage and common domain rules, reducing developer effort and improving coverage of edge cases.

\subsection{ Multi-level Contract Generation}

In this paper, we show how contracts at the method or class level help ensure the code quality generated by Vibe Coding. However, complex systems often require domain-specific or system-level contracts. Future research should explore formalizing contracts that capture domain-specific rules and interactions across modules, APIs, and services, including constraints on data flow, timing, and synchronization. For example, a Bank module might guarantee that a withdrawal operation never exceeds the account balance, while a bank in a different country may enforce different rules regarding balances or transaction limits. Learning these domain- or region-specific contracts may require retrieval-augmented generation (RAG) or fine-tuning of LLMs to encode the relevant rules or constraints. Multi-level contracts enhance system reliability and provide a rigorous framework for reasoning about correctness in large, modular systems.

\subsection{Runtime Contract Checker Development}

In this paper, we demonstrate the feasibility and benefits of contract-guided vibe coding. However, effectively enforcing these contracts at runtime remains a significant challenge, particularly across different programming languages and execution environments with varying levels of support for contract checking. A promising future direction is the systematic development of lightweight runtime checkers that automatically monitor contract satisfaction during execution. Such checkers can validate preconditions, postconditions, and invariants, detect violations early, and provide actionable feedback to both developers and LLMs. When integrated with the {\vibeC} paradigm, runtime checkers can further close the loop between contract specification, code generation, and execution, enabling adaptive refinement of contracts and prompts based on observed violations. Key research challenges include minimizing runtime overhead, handling partial or probabilistic contracts, and scaling contract monitoring to complex, stateful, and concurrent systems. Addressing these challenges would move contract-guided Vibe Coding toward stronger correctness guarantees and increased developer trust in LLM-generated software.

%% file: sections/conclusion.tex
\section{Conclusion}
\label{sec:conclusion}

We presented {\vibeC}, an enhancement to Vibe Coding that integrates Design by Contract principles into LLM-based code generation. By decomposing prompts into tasks, generating and validating contracts, and guiding code generation and testing, our approach prevents logic errors, enforces domain-specific constraints, and produces reliable code. {\vibeC}  lays the foundation for adaptive, verifiable, and trustworthy LLM-assisted software development, bridging the gap between high-level intent and robust, production-ready code.